\documentclass[a4paper,11pt]{article}
\usepackage{pos}
\usepackage{tikz}
\usetikzlibrary{positioning}
\usepackage{subcaption}

\title{Isospin Breaking Effects in the 2-Flavor Schwinger Model}

\author*{Nuha Chreim}
\author{Christian Hoelbling}
\author{Niklas Pielmeier}
\author{Lukas Varnhorst}

\affiliation[]{Department of Physics, University of Wuppertal, Gaußstraße 20, D-42119 Wuppertal, Germany}

\emailAdd{nuha.chreim@uni-wuppertal.de}

\abstract{The automatic fine-tuning of isospin breaking effects by conformal coalescence found by Georgi in the 2-flavor Schwinger model is studied. The analytical results obtained for the bosonic correlators are elaborated and the mass splitting parameter in leading order determined. Numerical investigation of meson mass splitting confirms the exponential suppression of symmetry breaking effects even for mass differences $\delta m$ near the fermion mass $m_f$.}

\FullConference{
  The 39th International Symposium on Lattice Field Theory\\
  8th-13th August, 2022\\
  Bonn, Germany 
}

\begin{document}
\maketitle

\newcommand*{\superscript}[1]{\ensuremath{^\textrm{{\scriptsize #1}}}}
\newcommand*{\subscript}[1]{\ensuremath{_\textrm{{\scriptsize #1}}}}
\newcommand*{\tinysuperscript}[1]{\ensuremath{^\textrm{{\tiny #1}}}}
\newcommand*{\tinysubscript}[1]{\ensuremath{_\textrm{{\tiny #1}}}}

\section{Introduction}
The Schwinger model \cite{Schwinger} exhibits many interesting features and its similarities to the QCD confined sector makes it a frequently studied toy model on the lattice. One unique and interesting feature is the bosonization of the massive model in the low energy sector, which enables the extraction of meson masses directly from the Lagrangian (see \cite{Sadzikowski_et_al,Delphenich}). In his fundamental paper from 1976 \cite{Coleman} Coleman discusses the massive Schwinger model and leaves three open questions concerning the 2-flavor case. Answers to some of these question have been provided prior to Georgi as can be found for example in \cite{Azcoiti}. Georgi provides answers to these questions through conformal coalescence in unparticle physics and shows that in the low energy sector of the Schwinger model, isospin breaking effects are automatically fine-tuned \cite{Georgi20}. \\
Large distance small energy dynamics in the massive Schwinger model have been studied prior by Smilga \cite{Smilga92,Smilga96} where the mass gap in the 2-flavor case and the analytic form of the bosonic correlator were determined in the bosonized theory.\\
The focus of this work lies on the analytical expansion and numerical verification of Georgis statement that isospin breaking effects are exponentially suppressed in the massive 2-flavor Schwinger model by powers of $\exp(-(\mu/m_f)^{2/3})$ where $\mu$ is the Schwinger mass and $m_f$ the fermion mass. The aim of this particular proceedings contribution is to give an explicit form of the mass splitting which was not provided in \cite{Georgi20}. Numerical results can be found in \cite{Niklas}.\\
For $N_f=2$ flavors, the bosonized Lagrangian in the 1+1D Schwinger model is given by
 \begin{equation}
          \mathcal{L}= \sum_{f = 1}^2 \frac{1}{2}  (\partial_{\mu} \Phi^{f})^2 -
       \frac{\mu^2}{4}  \left( \sum_{f = 1}^2 \Phi^f + \frac{\theta}{\sqrt{4 \pi}}
       \right)^2 + \sum_{f = 1}^2 cm_f^2 N_{m_f}  \left[ \cos \sqrt{4 \pi}
       \Phi^f \right] + const  
\end{equation}

 where $\Phi^f$ are pseudoscalar fields, $\mu = e\sqrt{\frac{2}{\pi}}$ is the Schwinger mass, $\theta$ the vacuum angle, $c={\frac{e^{\gamma}}{2{\pi} }}$ is a constant with $\gamma$ the Euler constant, $m_f$ the fermion mass for flavor $f$, and $N_{m_f}$ denotes normal-ordering with respect to the mass $m_f$.\\
 In the strong coupling limit for light quarks $(e \gg m_f)$ we can change the field variables by diagonalizing $\Phi^f$
        \begin{equation}
            \chi^a = O_f^a \Phi^f + \frac{\theta}{\sqrt{8 \pi}} \delta_1^a
            \label{chi2}
        \end{equation}
       
        using the orthogonal matrix
        \begin{equation}
          O_f = \frac{1}{\sqrt{2}} \begin{pmatrix}
             1 & 1\\
             1 & -1\\
            \end{pmatrix}.   
        \end{equation}
        
        Decoupling the heavy field $\chi^1$ that carries the mass $\sim\mu$ and renormal-ordering following Coleman \cite{Coleman} 
        \begin{equation}
            N_{m_f} \left[ \cos \left( - \frac{\theta}{2} + \sqrt{4 \pi} O_f^2 \chi^2
          \right) \right] = \left( \frac{M}{m_f} \right)^{\frac{1}{N_f}} N_M \left[
          \cos \left( - \frac{\theta}{2} + \sqrt{4 \pi} O_f^2 \chi^2 \right) \right],
        \end{equation}
        
        the resulting Lagrangian is that of the sine-Gordon theory with $\beta =
        \sqrt{2 \pi}$.
        \begin{equation}
           \mathcal{L}^{\text{light}} = \frac{1}{2}  (\partial_{\mu} \chi^2)^2 +
          \frac{1}{2 \pi} M^2 N_M  \left[ \cos \sqrt{2 \pi} \chi^2 \right]  
          \label{ssG}
        \end{equation}
        
        where the mass is given by
        \begin{equation}
            M = \left( e^{\gamma} \mu^{1 / 2}  \sqrt{m_1^2 + m_2^2 + 2 m_1 m_2 \cos
          \theta} \right)^{2 / 3} .
        \end{equation}
        
        Coleman identified three solutions: The soliton ($Q = 0$, $I_3 = +
        1$), the antisoliton ($Q = 0$, $I_3 = - 1$) and the lighter breather of the
        two breather solutions ($Q = 0$, $I_3 = 0$) that correspond to the pions
        $\pi^+, \pi^-$ and $\pi^0$ respectively. All three solutions are the lightest
        physical states of mass $M$ where $M$ is flavor dependent and
        the exponent of 2/3 is due to $N_f = 2$ \cite{Sadzikowski_et_al}.\\
        Following Georgi, the massless  composite operators of anomalous dimension 1/2 
        \begin{equation}
             O_f = \psi^{\ast}_{f 1} \psi_{f 2}, \quad  O^{\ast}_f = \psi^{\ast}_{f 2} \psi_{f 1}
             \label{cdo}
        \end{equation}
        and flavor $f = 1, 2$ have oppositely charged isospin components, $I_3 =+1, - 1$.
        Mixing these operators as
        \begin{equation}
            O_{\pm 1} = e^{i \theta / 2} O_1 \pm e^{- i \theta / 2} O_2^{\ast}  \qquad
           O_{\pm 1}^{\ast} = e^{- i \theta / 2} O^{\ast}_1 \pm e^{i \theta / 2} O_2
           \label{mo}
        \end{equation}
        all 2-point correlators vanish except for
        \begin{equation}
             \langle 0| T ( O_{\pm 1} (x) O^{\ast}_{\pm 1} (0)) | 0
          \rangle = \frac{\xi \mu}{2 \pi^2}  (e^{\kappa_0} \pm e^{- \kappa_0})
           \frac{1}{\sqrt{- x^2 + i \varepsilon}}
           \label{2pt}
        \end{equation}
    
        where
        \begin{equation}
            \kappa_0(x) = K_0  \left( \mu \sqrt{- x^2 + i \varepsilon} \right)
        \end{equation}
        with the Schwinger mass $\mu$ and $K_0$ is a modified Bessel function of the second kind.
        One can easily check that the $O_{- 1}$ correlator vanishes exponentially, while $O_1$ and $O^*_2$ in the $O_{+ 1}$ correlator become 
        identical in the long distance limit. This phenomenon is referred to as conformal coalescence by Georgi.\\
        The above formulation of operators and 2-point correlation functions come from the Sommerfield model which is massless fermions in 2D coupled to a massive vector field \cite{Georgi08,Sommerfield63}. It is an analog Banks-Zaks model \cite{Banks-Zaks}. 
        The Schwinger model is an asymptotic case of the Sommerfield model where the mass of the vector bosons go to zero \cite{Georgi08,Georgi19}.
        
\section{Degenerate Masses}
\label{dm}
We will now consider the case of stable bound isotriplets in the 2 flavor Schwinger model for $\theta = 0$. From the standard bosonization rules using eq.(\ref{chi2}) and $M \propto m_f^{\frac{3}{2}}$ we find the mass term in eq.(\ref{ssG}) in leading order 
        \begin{equation}
            \frac{1}{2 \pi} M^2 N_M  \left[ \cos \sqrt{2 \pi} \chi^2 \right] \propto m_f(\bar{\psi}_1 \psi_1 + \bar{\psi}_2 \psi_2).
        \end{equation}
        which in turn can be expressed using the composite operators as 
        \begin{align}
            \frac{1}{2 \pi} M^2 N_M  \left[ \cos \sqrt{2 \pi} \chi^2 \right] & \propto m_f(O_1+O_2)+h.c. 
        \end{align}
        Introducing a flavor degenerate mass term at low energies we use eq.(\ref{mo})
         \begin{equation}
             m_f (O_{+ 1} + O^{\ast}_{+ 1})
         \end{equation}
        
         and from the well known asymptotic behavior of $K_0$ and subsequently $\kappa_0$, i.e.
         \begin{equation}
             \kappa_0(x)\xrightarrow{x \rightarrow \infty} -i\sqrt{\frac{\pi}{2\mu x}}e^{-i(\mu x-\frac{\pi}{4})}
         \end{equation}
         in leading order we find that the exponential factors in eq.(\ref{2pt}) both asymptote to 1 as $x\rightarrow \infty$. Consequently the long distance behavior of the 2-point correlator of $O_{+1}(x)$ is given by 
         \begin{equation}
             \langle 0| T ( O_{+ 1} (x) O^{\ast}_{+ 1} (0))  | 0
           \rangle \xrightarrow{x \rightarrow \infty}  \frac{\xi \mu}{\pi^2} 
           \frac{1}{\sqrt{- x^2 + i \varepsilon}} = \frac{\xi \mu}{\pi^2}  \langle 0 T
           (O_{1 / 2} (x) O^{\ast}_{1 / 2} (0)) |0 \rangle
           \label{2pt+}
         \end{equation}
         where we have used the massless half dimensional conformal operator $O_{1 / 2}$ introduced by Georgi \cite{Georgi20,Georgi19} where the half dimension refers to the asymptotic anomalous dimension for large $\lvert x \rvert$ . The mass term may now be written as
        \begin{equation}
            \frac{\sqrt{\xi}}{\pi} m_f  \sqrt{\mu}  (O_{1 / 2} + O^{\ast}_{1 / 2})
        \end{equation}
         implying the relevant mass scale in the deep IR is given by $(m_f^2 \mu)^{\frac{1}{3}}$.
         Wick rotation to Euclidean space gives a complex spatial dimension \cite{Georgi22} and since we consider the long distance conformal sector 
         \begin{equation}
             (m_f^2\mu)^{-\frac{1}{3}}=\sqrt{-x^2}=ix.
         \end{equation}
        This mass scale is used in the following consideration to extract the isospin splitting term.
\section{Non-Degenerate Masses}
We now consider non-degenerate masses $m_f \pm \delta m$ with an isospin splitting term
         \begin{equation}
             \delta m (O_{- 1} + O^{\ast}_{- 1}).
         \end{equation}
         Looking once more at the asymptotic behavior of the exponential factors in eq.(\ref{2pt}) we find that in leading order 
         \begin{align}
             \frac{e^{\kappa_0}-e^{-\kappa_0}}{2} & \xrightarrow{x \rightarrow \infty} \frac{1}{2}(e^{-i\sqrt{\frac{\pi}{2\mu x}}e^{-i(\mu x-\frac{\pi}{4})}}-e^{i\sqrt{\frac{\pi}{2\mu x}}e^{-i(\mu x-\frac{\pi}{4})}}) \\
             & \xrightarrow{x \rightarrow \infty} -i\sqrt{\frac{\pi}{2\mu x}}e^{-i(\mu x-\frac{\pi}{4})}
         \end{align}
         and so the asymptotic behavior of the 2-point correlator of $O_{+1}(x)$ is  
         \begin{align}
             \langle 0| T ( O_{- 1} (x) O^{\ast}_{- 1} (0))  | 0
              \rangle \xrightarrow{x \rightarrow \infty}& -\frac{\xi \mu}{\pi^2} i\sqrt{\frac{\pi}{2\mu x}} e^{-i(\mu x-\frac{\pi}{4})} \langle 0| T (O_{1 / 2} (x) O^{\ast}_{1 / 2} (0)) |0 \rangle\\
               =&-i\xi \sqrt{\frac{\mu}{2 \pi^3x}} \sqrt{i}  e^{-i(\mu x)}  \langle 0| T (O_{1 / 2} (x) O^{\ast}_{1 / 2} (0)) |0 \rangle\\
               =&\xi \sqrt{\frac{1}{2 \pi^3}}  (m_f
               \mu^2)^{\frac{1}{3}} e^{- \left( \frac{\mu}{m_f} \right)^{\frac{2}{3}}} 
               \langle 0| T (O_{1 / 2} (x) O^{\ast}_{1 / 2} (0)) |0 \rangle
         \end{align}
       using the mass scale found in sect.\ref{dm} in the last step. The isospin splitting term is then
       \begin{equation}
           \delta m \sqrt{\xi \sqrt{\frac{1}{2 \pi^3}}  (m_f \mu^2)^{\frac{1}{3}} e^{-
           \left( \frac{\mu}{m_f} \right)^{\frac{2}{3}}}}  (O_{1 / 2} + O^{\ast}_{1 /
           2})
       \end{equation}
        and the isospin mass splitting scale is given by
        \begin{equation}
            \Delta M_s^3 = \delta m^2 m_f^{\frac{1}{3}} \mu^{\frac{2}{3}} e^{- \left(
           \frac{\mu}{m_f} \right)^{\frac{2}{3}}} .
        \end{equation}
        We note that the $O_{- 1} (x)$ correlator vanishes just one order more rapidly in $x^{- \frac{1}{2}}$ than the $O_{+ 1} (x)$ correlator and so the argument of $O_{- 1} (x)$ vanishing while $O_{+ 1} (x)$ goes to a conformal operator for $x\rightarrow \infty$ still holds.
        
         The overall mass term in the Lagrangian of half dimensional conformal operators is thus
         \begin{equation}
             \frac{\sqrt{\xi}}{\pi} m_f  \sqrt{\mu}  \left( 1 + \delta m\left(
          \frac{\pi}{2} \right)^{\frac{1}{4}} m_f^{- \frac{5}{6}} \mu^{- \frac{1}{6}}
          e^{- \frac{1}{2} \left( \frac{\mu}{m_f} \right)^{\frac{2}{3}}} \right)  (O_{1
          / 2} + O^{\ast}_{1 / 2})
         \end{equation}
        and knowing from the mass scale that $M_{\pi} \propto m_f^{\frac{2}{3}}$ we find the isospin breaking corrections to leading order in the $\delta m \xrightarrow{m_f \rightarrow 0} 0$ limit
        \begin{equation}
            M_{\pi} \propto m_f^{\frac{2}{3}}  \left( 1 + \frac{2}{3} \left(
          \frac{\pi}{2} \right)^{\frac{1}{4}} \frac{\delta m}{m_f^{\frac{5}{6}}
          \mu^{\frac{1}{6}}} e^{- \frac{1}{2} \left( \frac{\mu}{m_f}
          \right)^{\frac{2}{3}}} \right) .\label{splitt}
        \end{equation}
        Since the neutral pion operator in the nondegenerate case
        \begin{equation}
          O_{\pi^0} (x)  =  \frac{1}{2}  (\overline{\psi}_1 (x) \gamma_5 \psi_1 (x)
          - \overline{\psi}_2 (x) \gamma_5 \psi_2 (x))
        \end{equation}
        results in the propagator
        \begin{equation}
             \langle 0| T ( O_{\pi^0} (x) O^{\ast}_{\pi^0} (0)) |0 \rangle \propto \langle 0| T ( O_{+ 1} (x) O^{\ast}_{+ 1} (0))| 0 \rangle,
        \end{equation}
        the splitting term in eq.(\ref{splitt}) will only appear in the charged pion.
        
        For comparison, the  relation between the pion mass and the fermion mass in the degenerate two flavor case was found in \cite{Smilga96} and reads
        \begin{equation}
            M_{\pi}=2.008...m_f^{\frac{2}{3}}e^{\frac{1}{3}}.
        \end{equation}

\section{Lattice Observables}
In leading order the charged pion propagator does not feature isospin mass splitting. It consists of a connected term and is given by
\begin{equation}
	\langle O_{\pi^\pm}(x)\overline{O}_{\pi^\pm}(y)\rangle =-\text{tr}(D_u^{-1}(x,y)\gamma_5D_d^{-1}(y,x) \gamma_5). \label{con}
\end{equation}
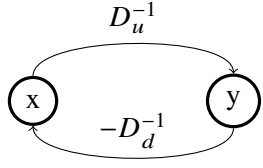
\begin{center}
  \begin{tikzpicture}	
    [
        roundnode/.style={circle, draw=black, fill=none, very thick, minimum size=3.5mm},
        squarednode/.style={rectangle, draw=red!60, fill=red!, very thick, minimum size=5mm},
        pointnode/.style={circle, very thick, minimum size=5mm}
    ]
	\node[draw=green!0]     (start)   {};
    \node[roundnode]     (upcircle)     [right=0.1cm of start]   {x};
    \node[pointnode]      (point)       [right=0.7cm of upcircle] {};
    \node[roundnode]        (downcircle)       [right=0.7cm of point] {y};
        				
        				
    \draw[->] (upcircle.north) .. controls +(up:5.5mm) and +(up:5.5mm) .. (downcircle.north) node[midway, above] {$D_u^{-1}$};
    \draw[->] (downcircle.south) .. controls +(down:5.5mm) and +(down:5.5mm) .. (upcircle.south) node[midway, above] {$-D_d^{-1}$};
\end{tikzpicture}  
\end{center}

The neutral pion propagator is given by 
\begin{align}	
    \langle O_{\pi^0}(x)\overline{O}_{\pi^0}(y)\rangle =&\frac{1}{2}  (-\text{tr}(D_u^{-1}(x,y)\gamma_5 D_u^{-1}(y,x)\gamma_5) +\text{tr}(D_u^{-1}(x,x)\gamma_5 )\text{tr}(D_u^{-1}(y,y)\gamma_5)\\
    &-\text{tr}(D_u^{-1}(x,x)\gamma_5)\text{tr}(D_d^{-1}(y,y)\gamma_5)
    + u \leftrightarrow d
    \end{align}
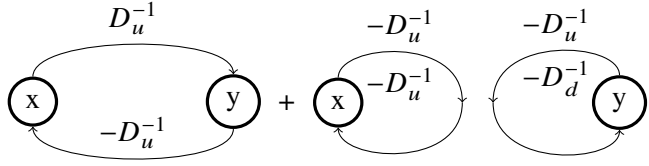
\begin{center}
    \begin{tikzpicture}	[
        	roundnode/.style={circle, draw=black, fill=none, very thick, minimum size=3.5mm},
        	squarednode/.style={rectangle, draw=red!60, fill=red!5, very thick, minimum size=5mm},
        	pointnode/.style={circle, very thick, minimum size=5mm}
        ]
    	\node[draw=green!0]     (start)   {};
    	\node[roundnode]     (upcircle)     [right=0.1cm of start]   {x};
    	\node[pointnode]      (point)       [right=0.7cm of upcircle] {};
    	\node[roundnode]        (downcircle)       [right=0.7cm of point] {y};
    				
    				
    	\draw[->] (upcircle.north) .. controls +(up:5.5mm) and +(up:5.5mm) .. (downcircle.north) node[midway, above] {$D_u^{-1}$};
    	\draw[->] (downcircle.south) .. controls +(down:5.5mm) and +(down:5.5mm) .. (upcircle.south) node[midway, above] {$-D_u^{-1}$};
    \end{tikzpicture}
    \begin{tikzpicture}[
    	roundnode/.style={circle, draw=black, fill=none, very thick, minimum size=3.5mm},
    	pointnode/.style={circle, very thick, minimum size=0.5mm}
    ]
    	\node[draw=green!0]     (start)   {$+$};
        \node[roundnode]        (upcircle)     [right=0.1cm of start]   {x};
    	\node[pointnode]      (point)       [right=1.3cm of upcircle] {};
        \node[roundnode]        (downcircle)       [right=1.3cm of point] {y};
    				
    	\draw[->] (upcircle.north) .. controls +(up:4.5mm) and +(up:9mm) .. (point.west) node[midway, above] {$-D_u^{-1}$} node[midway, below=0.07cm] {$-D_u^{-1}$};
    	\draw[->] (point.west) .. controls +(down:9mm) and +(down:4.5mm) .. (upcircle.south);
    	\draw[->] (downcircle.north) .. controls +(up:4.5mm) and +(up:9mm) .. (point.east) node[midway, above] {$-D_u^{-1}$} node[midway, below] {$-D_d^{-1}$};
    	\draw[->] (point.east) .. controls +(down:9mm) and +(down:4.5mm) .. (downcircle.south);
    	\end{tikzpicture}
\end{center}

which holds connected terms as given in eq.(\ref{con}) as well as disconnected terms that display the mass splitting. Since the connected term is free of isospin breaking effects in leading order, both connected terms in the charged and neutral pion propagator are equal. Ultimately the leading order neutral pion mass is given by
\begin{equation}
    M_{\pi^0}=M_{\pi^\pm}+\Delta M.
\end{equation}

To numerically investigate the exponential suppression of the splitting without further knowledge of the overall prefactor we consider the ratio 
\begin{equation}
    \frac{M_{\pi^\pm}+\Delta M}{M_{\pi^\pm}}=1+\frac{2}{3}\left(\frac{\pi}{2}\right)^{\frac{1}{4}}\frac{\delta m}{m_f^{\frac{5}{6}}\mu^{\frac{1}{6}}} e^{-\frac{1}{2}\left(\frac{\mu}{m_f}\right)^{\frac{2}{3}}}.
\end{equation}
from which only the splitting term is taken into account for the numerical results. Specifically we consider 
\begin{equation}
    \log \left(\underbrace{\frac{3}{2}\left(\frac{2}{\pi}\right)^{\frac{1}{4}}\frac{m_f^{\frac{5}{6}}\mu^{\frac{1}{6}}}{\delta m}}_{k}\frac{\Delta M}{M_{\pi^\pm}}\right)=-\frac{1}{2}\left(\frac{\mu}{m_f}\right)^{\frac{2}{3}}
    \label{meas}
\end{equation}
where k is kept constant. Numerical investigations of this behaviour are presented in a companion paper \cite{Niklas}. As detailed there, we find that the measured pion mass splitting does indeed follow the predicted exponential behavior, even when the mass splitting is relatively large. However, there seems to be a substantial deviation in the factor k, which is dependent on the splitting.

\section{Conclusion and Outlook}
We have studied and expanded on Georgi's analytical results of  isospin breaking effects by conformal coalescence in the 2-flavor Schwinger model \cite{Georgi20} and tested the exponential suppression of the mass splitting numerically on the lattice. The results confirm an exponential behavior in the mass splitting even for relatively large splittings. The offset to eq.(\ref{meas}) observed in the measurements indicates that there is a discrepancy in the factor $k$, which at the moment we do not understand. Hence, further analytic investigations, perhaps including a more systematic expansion procedure in $m_f$ and $\delta m$, will be necessary to obtain further insight.\\
Finally, we would like to point out that due to the different power counting schemes, the results obtained in this proceeding are not applicable to Georgi's latest publication where he considers small equal and opposite fermion masses \cite{Georgi22}.

\begin{flushleft}

\end{flushleft}
\end{document}